
\magnification=1200
\hsize=17 true cm
\vsize=22 true cm

\baselineskip=12pt

\def\[{\left\lbrack}
\def\]{\right\rbrack}
\def\hb{\hfill\break}
\def\({\left(}
\def\){\right)}
\def\{{\left \{}
\def\{{\right \}}
\def\ih{\'\i}
\def\hb{\hfill\break}

\hfill IF/UFRJ/93
\vskip 2 cm

\centerline{\bf REMARKS ON THE COLLECTIVE  QUANTIZATION }
\centerline{\bf OF THE SU(2) SKYRME MODEL}
\vskip 0.7 cm

\centerline{\it Jorge Ananias Neto \footnote*{\bf {Work supported
by CNPq. Brasilian Research Council}\hb \vbox { }\hskip .8 cm  e-mail:
$\,$IFT10030@UFRJ }}

\centerline{\it Instituto de F\ih sica,Universidade Federal do Rio de Janeiro}
\centerline{Rio de Janeiro, RJ 21945}

\vskip 0.7 cm

\noindent{\bf Abstract:} We point out the question of ordering
 momentum operator in the canonical \break  quantization of the SU(2) Skyrme
Model. Thus, we suggest a new definition for the  momentum operator
that may solve the infrared problem that appears when we try to
minimize the Quantum Hamiltonian.

\vskip 0.7 cm

\noindent PACS number(s): 03.65.Sq, 11.10.Ef, 12.40.Aa

\vfill\eject
\baselineskip=18pt

\noindent{\bf 1. Introduction }

\vskip .5 cm

\noindent  In the Skyrme Model baryons are treated as a soliton solution in a
non-Linear Sigma Model with an additional stabilizer Skyrme term
$^1$. The physical spectrum is obtained
performing the collective coordinate quantization.$\,$Using
the Nucleon and Delta masses as  input parameters,
we get the principal phenomenological results$^2$ . Although  most
of the static properties predicted by the Skyrme Model are in a good agreement
with the experimental results, certain values like, for example,
the pion decay constant,$\,F_\pi\, ,$
and the axial coupling constant, $\, g_A \,$,  present  large
 deviation from their experimental values.$\,$ However, we can overcome
 these problems attempting to
study with more detail the process of canonical quantization in the
rotational mode.$\,$ In previous works$^{3,4}$, some authors have pointed out
that the question of the quantization of  Skyrmions is a very  delicate one.
They mentioned that the Skyrmion quantization is a simple example of  quantum
mechanics on a curved space.
 \par This paper deals with the problem  of ordering that
appears in the definition of the canonical momentum when we try to
use  the constraint that is  present in the system.
Due to its simplicity it is not necessary to employ the Dirac
formalism of constraints$^{5,6}$.$\,$ We will observe that when we adopt
the correct definition for the momentum operator there is an
additional term in the Quantum Hamiltonian,$\,$ a result that has
been also obtained by the authors of ref.3,4  using another
calculate procedure.

\vskip .5 cm

\noindent {\bf 2. Quantization by Collective Coordinate Expansion}

\vskip .5 cm

\noindent Let us consider the classical static Lagrangian of the Skyrme Model
\vskip .2 cm

$$\eqalign{L=\int d^3r \[-{F^2_{\pi}\over16} \,\,\,\,Tr\left(\partial_i
U\partial_i  U^+\right)\,+
{1\over{32e^2}} \,Tr\left\lbrack U^+\partial_i U,
U^+\partial_j U\right\rbrack^2 \]\,\,,\cr}\eqno(1)$$
\vskip .3 cm

\noindent where $F_\pi$ is the pion decay constant, {\it$\, e \,$} is
a dimensionless parameter and U is an SU(2) matrix.

\noindent Performing  the collective semi-classical expansion$^2$,
substituting in (1) $\,\, U(r)\, by \,\,
U(r,t)=A(t)U(r)A^+(t) \, ,$
where  A is a SU(2) matrix, we obtain:
\vskip .2 cm

$$ L=-M+\lambda Tr\[\partial_0 A \partial_0 A^{-1}\] \,\,. \eqno(2) $$

\vskip .2 cm

\noindent In the last equation, M is the soliton mass which in the
 hedgehog representation
 for $ U,\,\, U=\exp{(i \tau . \hat r F(r))}\,$ ,$\,\,$ is given by

\vskip .2 cm

$$ M=4 \pi {F_\pi \over e} \int^{\infty}_0 x^2  {1\over 8} \[ F'^2 +
2{ \sin^2F \over x^2} \] + {1\over 2} {\sin^2F\over x^2}
\[ {\sin^2F \over x^2} + 2F'^2\] dx \,\, , \eqno(3)$$

\vskip .2 cm

\noindent where x is a dimensionless variable defined by $\,x=eF_{\pi}r \,
\,,$
and  $\,\,\lambda \,\, $ is called the inertia moment
written as
\vskip .1 cm

 $$\lambda={4\over6} \pi(1/e^3 F_{\pi}) \Lambda \,\, , \eqno(4)$$

\noindent with

$$ \Lambda=\int_o^\infty x^2 \sin^2F \[1+4\(F'^2+ \sin^2F\over{x^2} \)\] dx
\,\, . \eqno(5) $$

\vskip .9 cm

\noindent The SU(2) matrix A can be written as $ A=a_0+ia. \bf{\tau} $ ,
with the constraint

$$ \sum_{i=0} ^{i=3} a_i^2=0 \,\, . \eqno(6)$$

\vskip .5 cm

\noindent The Lagrangian (1) can be written as a function of
 the  $a^{\prime} s$ as:

\vskip .2 cm

$$ L=-M+2 \lambda \sum_{i=0} ^3 (\dot a _i)^2 \,\, . \eqno(7) $$

\vskip .2 cm

\noindent Introducing the conjugate momenta
$ \pi_i=\partial L/ \partial \dot a_i=4\lambda \dot a_i $ ,
 we can now rewrite the Hamiltonian as

\vskip .2 cm

$$ H=\pi_i \dot a_i-L=4\lambda \dot a_i \dot a_i -L=M+2 \lambda \dot a_i
\dot a_i
=M+{1\over 8 \lambda } \sum_i \pi_i^2 \,\, .  \eqno(8)  $$

\vskip .2 cm

\noindent Then, the standard canonical quantization is made where we replace
$\,\pi_i \,$ by $\, -i \partial/\partial a_i \,\, $ in (8) leading
to

\vskip .2 cm

$$ H=M+{1\over 8 \lambda } \sum_{i=0}^3 (-{\partial \over\partial a_i^2})
\,\, .  \eqno(9)  $$

\vskip .5 cm

\noindent Due to the constraint (6), the operator $  \sum_{i=0}^3
 (-{\partial \over\partial a_i^2}) $ is known as the Laplacian
 $ \nabla^2 $ on the three-sphere, with the eigenstates
 being traceless symmetric polynomials in the a$_i$. $\,$In order
to incorporate  relation (6)
it is more convenient to work  with  hypersphere coordinates defined by

\vskip .2 cm

$$\eqalign{ a_0 & = \cos W \cr
a_1 & = \sin W \cos \theta \cr
a_2 & = \sin W \sin \theta \cos \phi \cr
a_3 & = \sin W \sin \theta \sin \phi \,\,.}\eqno(10)$$

\vskip .4 cm

\noindent Then, the Laplacian written as a function of the hypersphere
 coordinates is given by

\vskip .2 cm

$$ \nabla^2 = {\partial^2 \over \partial W^2} + 2 {\cos W \over \sin W}
{\partial \over\partial W} + {1 \over \sin^2 W } { \partial^2 \over
\partial \theta^2} + {\cos \theta \over \sin^2 W \sin \theta} { \partial
 \over \partial \theta} + {1 \over \sin^2 W \sin^2 \theta}
 { \partial^2 \over \partial \phi^2} \,\,\,\, . \eqno (11)$$

\vskip .2 cm

\noindent Note that when $W=\pi/2 \,\, ,$  expression (11) reduces
to the classical Laplacian in spherical coordinates.
\noindent Thus, applying  expression (11) to the wave function
$ (a_0 + i a_1 )^l, \,$ we obtain

\vskip .2 cm

$$-\nabla^2(a_0 + i a_1 )^l=l(l+2) (a_0 + i a_1 )^l \,\,\,. \eqno(12) $$

\vskip .2 cm

\noindent If we wish to work with the  coordinates $\, a_i \,$ we must
be able to obtain an
expression for the canonical momentum $\,\pi_i \,$ .
\noindent We must remember that if the commutation relation

\vskip .2 cm

$$ \[a_ia_i,\pi_j\] = a_i\[a_i,\pi_j\]+\[a_i,\pi_j \]a_i \,\,   \eqno(13) $$

\vskip .2 cm

\noindent is valid, then the following relation must hold$^6$

\vskip .2 cm

$$ \[ a_i,\pi_j \] = \delta_{i,j} - a_ia_j \,\,\, . \eqno(14) $$

\vskip .2 cm

\noindent It is not difficult to see that  a possible expression for $\pi_j $,
 satisfying eq. (14), is given by$^5$

\vskip .2 cm

$$  \pi_j = {1\over i} \[\delta_{i,j} - a_ja_i\]\partial_i \,\,. \eqno(15)$$

\vskip .2 cm

\noindent Consequently,  $\pi_j\pi_j $ can be written as

\vskip .2 cm

$$ \pi_j\pi_j = -\partial_j\partial_j + 3 a_j \partial_j +
a_ia_j \partial_i\partial_j \,\, . \eqno(16) $$

\vskip .2 cm

\noindent Expression (16) is the three-sphere version of the Laplacian
 $\nabla^2$ written as a function of the coordinates a$_i$.
It should be noted that the
eigenvalues of the above equation are the same of those obtained
 using eq (12). At this point we must  mention
the problem of ordering that appears in the formula (15). As the physical
Hamiltonian must be Hermitian, the
usual choice for the operator momentum $\pi_j$, following the
prescription of Weyl ordering$^7$ is given by

\vskip .2 cm

$$  \pi_j = {1\over 2i} \[(\delta_{i,j} - a_ja_i)\partial_i
+\partial_i(\delta_{i,j}-a_ia_j)\] \,\,. \eqno(17)$$

\vskip .2 cm

\noindent If we substitute $\pi_j \,\,$ in eq.(16), we obtain
the following expression

\vskip .2 cm

$$ \pi_j\pi_j = -\partial_j\partial_j + 3 a_j \partial_j +
a_ia_j \partial_i\partial_j + {5\over4}\,\, . \eqno(18) $$

\noindent Comparing  expression (18) with (16) we see that an extra term
appears in the last equation. So, when we pay attention to the question
 of ordering in
the expression of the canonical momentum $\pi_j$ in the
coordinates $\, a_i \,$,
an additional term appears in the three-sphere \break Laplacian .Unfortunately,
if we want to improve the physical parameters predicted by the Skyrme Model,
 the signal of this extra term must be negative, as it was also shown
by A. Toda$^4$.

\par  As it was first point out by Bander and Hayot$^8$, if we observe
the asymptotic solution of the Euler-Lagrange equation that minimizes
the Quantum Hamiltonian (9)

\vskip .2 cm

$$ -{d^2F\over dx^2}-{2\over x} {dF\over dx}
+ {2\over x^2} F - k^2 F =0 \,\,\,\, , \eqno(19) $$

\vskip .1 cm

where $ k^2 $ is

\vskip .2cm

$$  k^2={3l(l+2) e^3 F_\pi  \over 8 \pi \( \int_o^\infty x^2 \sin^2F
 \[1+4\(F'^2+ \sin^2F\over{x^2} \)\] dx \)^2}
\,\, \,\,, \eqno(20) $$

\vskip .3 cm

\noindent then, we verify that F asymptotically  behaves as
 $ \,{\sin kr\over r}\, $ or
$ \, {\cos kr\over r} \,\,$  and the integral in the denominator of eq(20)
does not converge. The infrared problem is solved when we require
that the sign of the extra term is sufficiently negative in order
to modify  eq(19), which is now written as

\vskip .2 cm

$$ -{d^2F\over dx^2}-{2\over x} {dF\over dx}
+ {2\over x^2} F + k^2 F =0 \,\,\,\, . \eqno(21) $$

\vskip .3 cm

\noindent Studying the asymptotic  behaviour of F  we observe that
it behaves as $ \, {\exp{-kr}\over r} \,$
, and the integral in the denominator of eq(21)  converges.

In order to be able to deal with the problems that have been
presented by us in the previous lines we suggest a new definition for
the canonical operator momentum, which also satisfies
 the commutation  relation (14),

\vskip .2cm

$$  \pi_j = {1\over (1+\alpha)i} \[(\delta_{i,j} - a_ja_i)\partial_i
+{\alpha}\, \partial_i(\delta_{i,j}-a_ia_j)\] \,\,, \eqno(22)$$


\vfill\eject

\noindent where $\alpha \,\,$ is a free parameter. Consequently,
$\pi_j\pi_j \,\,$ is given by

\vskip .2cm

$$ \pi_j\pi_j = - \partial_j\partial_j
+3 a_j \partial_j + a_ja_i
\partial_j \partial_i -{5 \alpha \(2\alpha-3\)\over (1+\alpha)^2}
\,\,\, , \eqno (23)  $$

\vskip .2 cm

\noindent and the eigenvalues of the extended Quantum Hamiltonian are
given by
\vskip .2 cm

$$ E=M+{1\over 8 \lambda } \[  l(l+2) -
{5 \alpha (2 \alpha - 3) \over (1+\alpha)^2} \] \,\,\,. \eqno(24) $$

\vskip .4 cm

\noindent In the above equation we observe that when
$ \,\alpha > {3\over 2} \,\,,$ the extra term
is negative. For the nucleon state, l=1, we verify that for
$ \,\, \alpha > {21+5\sqrt 21 \over 14} \,$ or $ \,\,\alpha < {21-5\sqrt
21 \over 14}\,\,$ there is no infrared problem in the quantum
Hamiltonian, as we have remarked in (19). Now it is possible
to search to a solution F(x) that minimizes the total Quantum Hamiltonian
written in (24).

\vskip 1 cm

\noindent {\bf3. Conclusion}

\vskip .5 cm

\noindent We have shown that with the definition of the canonical momentum,
$\,\pi_i\,$, which rules the Weyl prescription of ordering$^7 \,$, there
is an additional term in the usual Skyrmion quantization.$\,$ It is
possible to redefine the expression of the canonical momentum,$\,$
which also satisfies the commutation relation of a particle in the
three-sphere,$\,$ with the purpose of removing the infrared problem.$\,$
We hope that, with the use of the quantum variational
solution,$\, F(r)\,$, one can be able to obtain an improvement of the
physical parameters.$\,$ The behaviour of these solutions$^9$ and the extension
of this analysis to the SU(N) Collective Quantization
, in particular in the case of the SU(3) Skyrmions$^{10} \,$
will be  objects of forthcoming papers$^{11,12}$.

\vskip 1 cm

\par We would like to thank M.G. do Amaral, P. Gaete and S.M.de Souza
 for useful discussions.
\vskip 1 cm

\vfill\eject

\noindent {\bf References}
\vskip .4 cm

\noindent 1.T.H.R. Skyrme, Proc. Roy. Soc. A260 (1961) 127.\hb
2.G.S. Adkins, C.R. Nappi and E. Witten, Nucl. Phys. B228 (1983) 552.\hb
3.K.Fujii, K.I.Sato, N.Toyota and A.P. Kobushkin, Phys. Rev. Lett.58, 7 (1987)
\hb  \vbox { }\hskip .2 cm and Phys.Rev.D35,6 (1987) 1896.\hb
4.A.Toda, Prog.Theor.Phys. 84(1990) 324.\hb
5.H.J. Schnitzer, Nucl.Phys.B261 (1985) 546. \hb
6.H. Verschelde and H. Verbeke, Nucl. Phys.A500 (1989) 573. \hb
\vbox { }\hskip .2 cm N. Ogawa, K. Fujii and A. Kobushkin, Prog. Theor. Phys.
83 (1990) 894.\hb
7.The same procedure was performed by A.Toda(ref.4) in another context.\hb
\vbox { } \hskip .2cm T.D. Lee, Particle Physics and Introduction
 to Field Theory p.476.(Harwood Academic, New York,1981)\hb
8.M. Bander and F. Hayot, Phys.Rev.D30,8 (1984) 1837.\hb
9.E. Ferreira and J. Ananias Neto, J.Math.Phys.33(3) (1992) 1185. \hb
10.H. Yabu and K. Ando Nucl.Phys.B301 (1988) 601.\hb
\vbox { } \hskip .2cmC. Callan and I. Klebanov, Nucl.Phys.B262 (1985) 365. \hb
11.Jorge Ananias Neto, work in progress.\hb
12.Jorge Ananias Neto, work in progress.

\bye